\documentclass{article}

\usepackage{graphicx}

\usepackage{amsmath,latexsym}

\begin{document}

\setcounter{page}{1}

\pagestyle{plain}

\begin{center}
\Large{\bf $\alpha$-Attractor and Reheating in a Model with Non-Canonical Scalar Fields}\\
\small \vspace{1cm} {\bf Narges
Rashidi\footnote{n.rashidi@umz.ac.ir}}\quad and \quad {\bf
Kourosh Nozari\footnote{knozari@umz.ac.ir(Corresponding Author)}} \\
\vspace{0.5cm} Department of Physics, Faculty of Basic
Sciences,
University of Mazandaran,\\
P. O. Box 47416-95447, Babolsar, Iran\\
and\\
Research Institute for Astronomy and
Astrophysics of Maragha (RIAAM),\\
P. O. Box 55134-441, Maragha, Iran
\end{center}

\begin{abstract}
We consider two non-canonical scalar fields (tachyon and DBI) with
E-model type of the potential. We study cosmological inflation in
these models to find possible $\alpha$-attractors. We show that
similar to the canonical scalar field case, in both tachyon and DBI
models there is a value of the scalar spectral index in small
$\alpha$ limit which is just a function of the e-folds number.
However, the value of $n_{s}$ in DBI model is somewhat different
from the other ones. We also compare the results with Planck2015 TT,
TE, EE+lowP data. The reheating phase after inflation is studied in
these models which gives some more constraints on the model's parameters.
\end{abstract}
\newpage

\section{Introduction}\label{se1}

It is now accepted that the physics of the early universe can be
explained by a testable paradigm named cosmological inflation. The
simplest realization of the inflation is a model with a
canonically-normalized single scalar field which its nearly flat
potential dominates the energy density of the universe. In this
model, the dominant mode of the primordial density perturbations
(seeded by the quantum fluctuations of the scalar field during the
inflation era) is predicted to be almost adiabatic and scale
invariant and has Gaussian
distribution~\cite{Gut81,Lin82,Alb82,Lin90,Lid00a,Lid97,Rio02,Lyt09,Mal03}.
However, there is a possibility that inflation may be driven by a
single field with non-canonical kinetic energy. Usually, the
non-canonical inflation models are referred to as ``k-inflation''.
These models predict that the primordial density perturbations are
somehow scale dependent (which is mildly supported by the Planck2015
released data~\cite{Ade15a,Ade15b}) and have non-Gaussian
distribution. Among the k-inflation models, we can mention the DBI
and Tachyonic models. In the DBI (Dirac-Born-Infeld) model, the D3
brane moves in a (usually $AdS_{5}$) throat region of a warped
compactified space and its radial coordinate identifies the inflaton
field~\cite{Sil04,Ali04}. In this model the action involves a
non-canonical kinetic term. Also there is a function of the scalar
field besides the potential in the action. This function is related
to the local geometry of the compact manifold through it the D3
brane traverses. Tachyon field also, is associated to the D-branes
in string theory~\cite{Sen99,Sen02a,Sen02b}. This field can be
responsible for early time inflation in the history of the Universe,
as well as, the late time accelerating expansion. Authors have
studied some aspects of the tachyon and DBI models in
Refs.~\cite{Sam02,Fein02,Gib02,Noz13a,Noz14,Ota13,Hua06,Che07,Noz13b,Miz10,Spa07}

The ``cosmological attractor'' in inflation models is the idea which
has attracted much attention recently. There are several models
incorporating the idea of cosmological attractors which among them
we refer to conformal attractors~\cite{Kal13a,Kal13b} and
$\alpha$-attractors models~\cite{Kai14,Fer13,Kal13c,Kal14}.
In~\cite{Cec14,Kal14b,Lin15,Jos15a,Jos15b,Kal16,Sha16,Odi16} one can
find more details on the issue of $\alpha$-attractors. The important
issue in the conformal attractor model is that in the large e-folds
number ($N$), it has the universal prediction as
$n_{s}=1-\frac{2}{N}$ and $r=\frac{12}{N^{2}}$. The
$\alpha$-attractor models have two types called E-model and T-model
according to the adopted potentials. The potential characterizing
the E-model is given by
\begin{equation}
\label{eq1}
V=V_{0}\Big[1-\exp\big(-\sqrt{\frac{2\kappa^{2}}{3\alpha}}\phi\big)\Big]^{2n}\,,
\end{equation}
and the potential characterizing the T-model is defined as
\begin{equation}
\label{eq2}V=V_{0}\tanh^{2n}\Big(\frac{\kappa\phi}{\sqrt{6\alpha}}\Big)\,,
\end{equation}
with $V_{0}$, $n$ and $\alpha$ being some free parameters. It is
shown that a canonical single field $\alpha$-attractor model, in the
small $\alpha$ limit predicts $n_{s}=1-\frac{2}{N}$ and
$r=\frac{12\alpha}{N^{2}}$. As we see, in small $\alpha$ and large
$N$ limit, the prediction of the scalar spectral index in the
$\alpha$-attractor models is the same as the prediction in the
conformal attractor models. In this limit, the tensor-to-scalar
ratio in $\alpha$-attractor models is a function of $\alpha$,
whereas, it is independent of $\alpha$ in the conformal attractor
models.

In the study of cosmological inflation, the reheating process after
the end of inflation is an important issue. The universe inflates as
long as the potential is sufficiently flat and the slow-roll
conditions $\eta,\epsilon\ll 1$ are satisfied. The inflaton rolls
into the minimum of its potential, then as soon as the slow-roll
conditions break down and inflation ends it starts to oscillate
about the minimum. According to the simple canonical reheating
scenario, when inflaton oscillates, it loses energy and by passing
the processes which include the physics of particle creation and
non-equilibrium phenomena, decays into the plasma of the
relativistic particles corresponding to the radiation-dominated
Universe~\cite{Ab82,Do82,Al82}. Nevertheless, some authors have
proposed other complicated scenarios of reheating including the
non-perturbative processes. The instant preheating~\cite{Fel99}, the
parametric resonance decay~\cite{Ko94,Tr90,Ko97} and tachyonic
instability~\cite{Gr97,Sh06,Du06,Ab10,Fel01a,Fel01b} are the
examples among the non-perturbative reheating scenarios. Some
important parameters, characterizing the reheating epoch, are the
e-folds number during reheating ($N_{rh}$) and the reheating
temperature ($T_{rh}$). Exploring these parameters during inflation
models helps us to find some more constraints on the models
parameters~\cite{Dai14,Un15,Co15,Cai15,Ue16,Noz16b}. Another useful
parameter to study the reheating phase is the effective equation of
state parameter during reheating ($\omega_{eff}$). The value of the
effective equation of state parameter for a massive inflaton can be
$-1$ (if the potential dominates the energy density) and $+1$ (if
the kinetic term dominates the energy density). Regarding to this
fact that the value of $\omega_{eff}$ at the end of the inflation
epoch is $-\frac{1}{3}$ and its value at the beginning of the
radiation dominated universe is $\frac{1}{3}$, it seems logical to
assume the effective equation of state parameter during the
reheating epoch in the range
$-\frac{1}{3}\leq\omega_{eff}\leq\frac{1}{3}$. The frequency of the
oscillations of the massive inflaton is very larger than the
expansion rate at the initial epoch of the reheating, leading to the
vanishing averaged effective pressure. In this respect, at the
beginning of the reheating epoch the effective equation of state
parameter can be considered to be zero, effectively corresponding to
the equation of state parameter of the dust matter. After that, when
the inflaton oscillates and decays into other particles, the value
of $\omega_{eff}$ increases with time and reaches $\frac{1}{3}$,
when the radiation dominated era begins. In this regard, this
parameter also gives some constraints on the model's parameters. See
also Ref.~\cite{Am15} for a review on reheating.

\begin{figure}[ht]
\begin{center}
\includegraphics[scale=0.6]{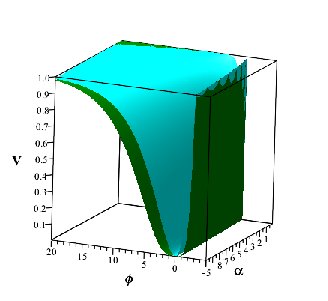}
\end{center}
\caption{\small {Evolution of the E-model type of
potential with $n=1$ (cyan) and $n=2$ (green). For all values of
$\alpha$, the potential at large positive values of $\phi$ is nearly
flat.}}
\label{fig1}
\end{figure}

In this paper we consider two inflation models with non-canonical
kinetic term: the Tachyon and DBI models. As is clarified
in Ref.~\cite{Ali04}, $f(\phi)$ in DBI model is the warp factor of the
AdS throat which for $AdS_5$ throat it is equal to
$\frac{\lambda}{\phi^{4}}$. Also, if we consider the $AdS_5 \times X$
geometry, the potential of a DBI field would be quartic. For an
approximate AdS throat, there would be a massive scalar field with
quadratic potential. On the other hand, in Ref.~\cite{Tsuj13} it has been
shown that with $f\sim e^{\lambda\phi}$ and $V\sim e^{-\lambda\phi}$
(with $\lambda$ to be a constant) we can get the Lagrangian of the DBI
model. Also, the authors of Ref.~\cite{Kin08} have obtained the mentioned
functions in the DBI inflation model. In Ref.~\cite{Sen02b} it has been
demonstrated that potential of the tachyon model is proportional
to $e^{-\beta\phi}$, where $\beta$ is a constant. Also, some authors
have studied tachyon cosmology with power law
potential (for instance ~\cite{Noz13a,Noz14,Li14}) and inverse power law
potential~\cite{Zha10}. Our motivation in this work was two-folds: firstly we have tried to combine,
two successful ingredients of inflationary model-building,
that is, non-canonical kinetic terms that facilitate
the slow-roll inflation and alpha-attractor potentials
that provide robust predictions with the hope to shed
more light on these issues. Secondly, this model provides
a framework that some of the previous studies are
special subclass of the solutions presented here. In this regard, by adopting an E-model
potential in both the tachyon and DBI model (and also E-model
$f^{-1}(\phi)$ in the DBI model), we are able to cover the mentioned types of the potentials.
For instance, in large $\alpha$ limit, we have power law
inflation. In small $\alpha$ limit (but not $\alpha\rightarrow 0$)
we get the inverse exponential potential. We have similar situation for
$f(\phi)$. In large $\alpha$ limit, we have $f^{-1}(\phi)\sim
\phi^{2n}$. In small $\alpha$ limit we reach an exponential type of
$f(\phi)$. In this regard, to study cosmological dynamics of tachyon
and DBI models we adopt E-model type of potential with $n=1$ and $n=2$. As figure 1
shows, this potential at large positive values of the scalar field
is nearly flat. By assuming this potential, in section II we obtain
the slow roll parameters, the scalar spectral index and the
tensor-to-scalar ratio in both non-canonical models. We show that
the tachyon inflation model, at large $N$ and small $\alpha$,
predicts the same scalar spectral index and tensor-to-scalar ratio
as the ones predicted in the canonical single field inflation.
However, the DBI model predicts the scalar spectral index somewhat
different. We also study the evolution of the tensor-to-scalar ratio
versus the scalar spectral index in the background of Planck2015 TT,
TE, EE+lowP data. As we shall see, the DBI model with E-model
potential and for both $n=1$ and $n=2$, does not lie within the
95$\%$ confidence region of the $n_{s}-r$ plane. In section III, we
study the reheating phase in the tachyon and DBI models. We obtain
the e-folds number, temperature and effective equation of state
during reheating. By comparing with observational data, we
constraint the model's parameters.

\section{Inflation}\label{se2}
The general action for an inflation model driven by an arbitrary
single scalar field is given by
\begin{eqnarray}
\label{eq3} S=\int
d^{4}x\sqrt{-g}\Bigg[\frac{1}{2\kappa^{2}}R+P(X,\phi) \Bigg],
\end{eqnarray}
where, $R$ is the Ricci scalar and the kinetic energy of the scalar
field ($\phi$) is defined as
$X=-\frac{1}{2}\partial_{\nu}\phi\,\partial^{\nu}\phi$. To study the
cosmological dynamics, the term $P(X,\phi)$ should be specified.
This term for the tachyon ($tch$) and DBI models is defined as
\begin{equation}
\label{eq4}P_{tch}(X,\phi)=-V(\phi)\sqrt{1-2X}\,,
\end{equation}
and
\begin{equation}
\label{eq5}P_{DBI}(X,\phi)=-f^{-1}(\phi)\sqrt{1-2f(\phi)X}-V(\phi)\,,
\end{equation}
respectively. To proceed, we consider each model separately and
study its dynamics.

\subsection{Inflation in the tachyon model with E-model potential}\label{se2-1}

In a spatially flat FRW metric, the action (\ref{eq3}) with
$P(X,\phi)$ defined in (\ref{eq4}) leads to the following Friedmann
equation
\begin{equation}
\label{eq6}H^{2}=\frac{\kappa^{2}}{3}\frac{V}{\sqrt{1-\dot{\phi}^{2}}}\,,
\end{equation}
where a dot denotes cosmic time derivative of the parameter. By
varying the action (\ref{eq3}), by $P(X,\phi)$ defined in
(\ref{eq4}), with respect to the scalar field, the following
equation of motion is obtained
\begin{equation}
\label{eq7}\frac{\ddot{\phi}}{1-\dot{\phi}^{2}}+3H\dot{\phi}+\frac{V'}{V}=0\,,
\end{equation}
where a prime shows derivative with respect to the tachyon field. To
have inflation phase, the slow roll parameters, defined as
$\epsilon\equiv-\frac{\dot{H}}{H^{2}}$ and
$\eta=-\frac{1}{H}\frac{\ddot{H}}{\dot{H}}$, should satisfy the
conditions $\epsilon \ll 1$ and $\eta \ll 1$ (meaning that
$\dot{\phi}^{2}\ll1$ and $\ddot{\phi}\ll 3H\dot{\phi}$). In this
regard we obtain
\begin{equation}
\label{eq8}\epsilon=\frac{1}{2\kappa^{2}}\frac{V'^{2}}{V^{3}}\,,
\end{equation}
and
\begin{equation}
\label{eq9}\eta=\frac{1}{\kappa^{2}}\left[\frac{V''}{V^{2}}-\frac{1}{2}\frac{V'^{2}}{V^{3}}\right]\,,
\end{equation}
which in the inflationary era are much smaller than unity and when
one of them reaches unity the inflation ends. By using the
definition of the e-folds number during inflation as
\begin{equation}
\label{eq10} N=\int_{t_{hc}}^{t_{e}} H dt\,,
\end{equation}
with $t_{hc}$ and $t_{e}$ being the time of the horizon crossing and
end of inflation respectively, we get the following expression
\begin{equation}
\label{eq11} N \simeq \int_{\phi_{hc}}^{\phi_{e}}
\frac{-\kappa^{2}V^{2} }{V'} d\phi\,.
\end{equation}

To obtain the perturbation parameters (the scalar spectral index and
tensor-to-scalar ratio), we use the power spectrum defined as
\begin{equation}
\label{eq12}{\cal{A}}_{s}=\frac{H^{2}}{8\pi^{2}{\cal{W}}_{s}c_{s}^{3}}\,,
\end{equation}
where
\begin{equation}
\label{eq13} {\cal{W}}_{s}=\frac {\dot{\phi}^{2}\,V}{2 \big(
1-\dot{\phi}^{2}\big) ^{\frac{3}{2}}H^{2}}\,,
\end{equation}
and the sound speed is given by
\begin{equation}
\label{eq14} c_{s}=\sqrt{1-\dot{\phi}^{2}}\,.
\end{equation}
The parameters ${\cal{A}}_{s}$ and ${\cal{W}}_{s}$ are
evaluated at the horizon crossing time. The
scalar spectral index is obtained by using the power spectrum as
follows
\begin{eqnarray}
\label{eq15} n_{s}-1=\frac{d \ln {\cal{A}}_{s}}{d \ln
k}\Bigg|_{c_{s}k=aH}\,,
\end{eqnarray}
which gives
\begin{eqnarray}
\label{eq16} n_{s}=1-6\epsilon+2\eta\,.
\end{eqnarray}
Also, the tensor-to-scalar ratio in this setup is given by
\begin{eqnarray}
\label{eq17} r=16c_{s}\epsilon\,.
\end{eqnarray}
To see more details about obtaining equations
(\ref{eq8})-(\ref{eq17}) see
Refs.~\cite{Noz13a,Fel11a,Fel11b,Che08}.

Now, we study the tachyon model with E-model potential defined in
(\ref{eq1}). First, we seek for the scalar spectral index and
tensor-to-scalar ratio in the large $N$ and small $\alpha$ limit. In
this limit, we can rewrite the E-model potential as
\begin{equation}
\label{eq18}V=V_{0}\Big[1-2n\exp\big(-\sqrt{\frac{2\kappa^{2}}{3\alpha}}\phi\big)\Big]\,.
\end{equation}
With this potential, the slow-roll parameter $\epsilon$ takes the
following form
\begin{eqnarray}
\label{eq19}\epsilon=\frac{4}{3}\,{n}^{2} \frac{\left( {{\rm
e}^{-\frac{\sqrt {6}}{3}\sqrt {{\frac {{\kappa}^{
2}}{\alpha}}}\phi}} \right) ^{2}}{\left( 1-2\,n{ {\rm
e}^{-\frac{\sqrt {6}}{3}\sqrt {{\frac {{\kappa}^{2}}{\alpha}}}\phi}}
\right) ^{3}}\,{\alpha}^{-1}\,.
\end{eqnarray}
The value of $\epsilon$ at horizon crossing is obtained by setting
$\phi=\phi_{hc}$, where $\phi_{hc}$ is found from equation
(\ref{eq11}) (in which we assume $\phi_{e}\ll \phi_{hc}$). By
substituting the obtained $\phi_{hc}$ and considering that the
expression ${{\rm e}^{-\frac{\sqrt {6}}{3}\sqrt {{\frac {{\kappa}^{
2}}{\alpha}}}\phi}}$ in the considered limit is very small, we
obtain
\begin{eqnarray}
\label{eq20}\epsilon=\frac{3}{4}\frac{\alpha}{N^{2}}\,.
\end{eqnarray}
The above equation by using the definition (\ref{eq17}) leads to
\begin{eqnarray}
\label{eq21}r=\frac{12\alpha}{N^{2}}\,,
\end{eqnarray}
which is exactly the same as the predicted tensor-to-scalar ratio in
the large $N$ and small $\alpha$ limit obtained in the canonical
single field inflation. Similarly, for the scalar spectral index, by
using $\phi_{hc}$ and equation (\ref{eq16}) we find
\begin{eqnarray}
\label{eq22}n_{s}=1-\frac{6\,\alpha}{{N}^{2} \left(
1-\frac{3}{2}\,{\frac {\alpha}{N}} \right) ^{3}}- \frac{2}{{N}
\left( 1-\frac{3}{2}\,{\frac {\alpha}{N}} \right) ^{2}}\,.
\end{eqnarray}
The above expression, in the large $N$ and small $\alpha$ limit
becomes
\begin{eqnarray}
\label{eq23}n_{s}=1-\frac{2}{N}.
\end{eqnarray}
In this limit, the scalar spectral index in tachyon model is also
the same as the one predicted in the canonical scalar field model.

On the other hand, if we consider $\alpha\rightarrow \infty$, the
E-model potential tends to $\phi^{2n}$ leading to
\begin{equation}
\label{eq24}\epsilon={\frac
{2{n}^{2}}{{\phi}^{2n+2}{\kappa}^{2}}}\,,
\end{equation}
and
\begin{eqnarray}
\label{eq25}n_{s}=1-4\,{\frac {n \left( 2\,n+1 \right)
}{{\phi}^{2n+2}{\kappa}^{2}}}\,.
\end{eqnarray}
To numerical study of the perturbation parameters $r$ and $n_{s}$
and comparing them with observational data, we use equations
(\ref{eq17}) (where $\epsilon$ is given by equation (\ref{eq8}) with
potential (\ref{eq1})) and (\ref{eq22}). The results are shown in
figure 2. As this figure shows, for both $n=1$ and $n=2$ cases, the
scalar spectral index and tensor-to-scalar ratio in
$\alpha\rightarrow 0$ limit, tend to $n_{s}=0.96$ and $r=0$ (for
$N=50$) and $n_{s}=0.966$ and $r=0$ (for $N=60$). In large $\alpha$
limit, the model reaches the tachyon inflation with power law
potential. For $n=1$, in the large $\alpha$ limit, we get $\phi^{2}$
tachyon inflation and for $n=2$, we get $\phi^{4}$ tachyon
inflation. Note that, the tachyon model with E-model potential (and
with both $n=1,2$ and $N=50,60$) for all values of $\alpha$ is
consistent with the Planck2015 TT, TE, EE+lowP data.

\begin{figure}[ht]
\begin{center}
\includegraphics[scale=0.6]{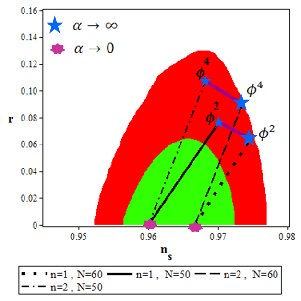}
\end{center}
\caption{\small {Tensor-to-scalar ratio versus the
scalar spectral index for a tachyon model with the E-model
potential. The smaller blue stars are corresponding to the tachyon
inflation with $\phi^{2n}$ potential and $N=50$ and the larger blue
stars are corresponding to the tachyon inflation with $\phi^{2n}$
potential and $N=60$.}}
\label{fig2}
\end{figure}

\subsection{Inflation in the DBI model with E-model potential}\label{se2-2}
Now, we study inflation in the DBI model. The action (\ref{eq3})
with $P(X,\phi)$ defined in (\ref{eq5}), gives the following
Friedmann equation
\begin{equation}
\label{eq26}H^{2}=\frac{\kappa^{2}}{3}\Bigg[\frac{f^{-1}}{\sqrt{1-f\,\dot{\phi}^{2}}}+V\Bigg]\,.
\end{equation}
Varying the action (\ref{eq3}), by $P(X,\phi)$ defined in
(\ref{eq5}), with respect to $\phi$ leads to the following equation
of motion
\begin{equation}
\label{eq27}\frac{\ddot{\phi}}{(1-f\dot{\phi}^{2})^{\frac{3}{2}}}+\frac{3H\dot{\phi}}{(1-f\dot{\phi}^{2})^{\frac{1}{2}}}+V'=-\frac{f'}{f^{2}}
\Bigg[\frac{3f\dot{\phi}^{2}-2}{2(1-f\dot{\phi}^{2})^{\frac{1}{2}}}\Bigg].
\end{equation}
Inflation occurs when the conditions $\epsilon \ll 1$ and $\eta \ll
1$ (corresponding to $f\dot{\phi}^{2}\ll1$ and $\ddot{\phi}\ll
3H\dot{\phi}$) are satisfied, where
\begin{equation}
\label{eq28}\epsilon={\frac {{f}^{2}{{\it V'}}^{2}}{2{\kappa}^{2}
\left( Vf+1 \right) ^ {2}}}-{\frac {{\it V'}\,{\it f'}}{{\kappa}^{2}
\left( Vf+1 \right) ^{2 }}}+{\frac {{{\it
f'}}^{2}}{2{f}^{2}{\kappa}^{2} \left( Vf+1 \right) ^{2}}} \,,
\end{equation}
and
\begin{equation}
\label{eq29}\eta=-{\kappa}^ {-2} \left(  \frac{\left( 2\,{\it
V''}-2\,{\frac {{\it f''}}{{f}^{2}}} \right)}{ \left( V+{f}^{-1}
\right) }-\frac{ \left( {\it V'}-{\frac {{\it f'}}{{f }^{2}}}
\right) ^{2}}{ \left( V+{f}^{-1} \right)} \right)  \,.
\end{equation}
The e-folds number during inflation in DBI model is given by
\begin{equation}
\label{eq30} N\simeq \int_{\phi_{hc}}^{\phi_{e}}
\frac{\kappa^{2}(V+f^{-1}) }{-V'+f'f^{-2}} d\phi\,.
\end{equation}
The power spectrum in this model is given by equation (\ref{eq13})
with new definition of ${\cal{W}}$ and $c_{s}$ as
\begin{equation}
\label{eq31} {\cal{W}}_{s}={\frac {\dot{\phi}^{2}}{ 2\left(
1-f\,\dot{\phi}^{2} \right) ^{3/2}H^{2}}}\,,
\end{equation}
and
\begin{equation}
\label{eq32} c_{s}=\sqrt{1-f\,\dot{\phi}^{2}}\,.
\end{equation}
The scalar spectral index and the tensor-to-scalar ratio are given
by equations (\ref{eq16}) and (\ref{eq17}) with the slow-roll
parameters defined in (\ref{eq28}) and (\ref{eq29}).

Similar to the tachyon model, we study the DBI model with E-model
potential defined in (\ref{eq1}) and
\begin{equation}
\label{eq33}
f=f_{0}\Big[1-\exp\big(-\sqrt{\frac{2\kappa^{2}}{3\alpha}}\phi\big)\Big]^{-2n}\,.
\end{equation}
To explore the scalar spectral index and tensor-to-scalar ratio in
large $N$ and small $\alpha$ limit, we use the potential
(\ref{eq18}) and
\begin{equation}
\label{eq34}f=f_{0}\Big[1+2n\exp\big(-\sqrt{\frac{2\kappa^{2}}{3\alpha}}\phi\big)\Big]\,,
\end{equation}
which is written in this limit. With this potential, the slow-roll
parameter $\epsilon$ in DBI model is given by the following
expression
\begin{eqnarray}
\label{eq35}\epsilon=\frac{4}{3}{\alpha}^{-1}\,{n}^{2}{{\rm
e}^{-\frac{2}{3}\,{\frac {\sqrt {6}\kappa\,\phi}{\sqrt {\alpha}}}}}
\bigg[ 4\,{{\rm e}^{-\frac{4}{3}\,{\frac {\sqrt {6}\kappa\,\phi}{
\sqrt {\alpha}}}}}{n}^{4} +8\,{{\rm e}^{-{\frac {\sqrt
{6}\kappa\,\phi} {\sqrt {\alpha}}}}}{n}^{3}+8\,{n}^{2}{{\rm
e}^{-\frac{2}{3}\,{\frac {\sqrt {6} \kappa\,\phi}{\sqrt
{\alpha}}}}}+4\,n{{\rm e}^{-\frac{1}{3}\,{\frac {\sqrt
{6} \kappa\,\phi}{\sqrt {\alpha}}}}}+1 \bigg]\times\nonumber\\
 \left[\Big( 1+2\,n {{\rm e}^{-\frac{1}{3}\,{\frac {\sqrt
{6}\kappa\,\phi}{\sqrt {\alpha}}}}} \Big) \Big( 2\,{n}^{2}{{\rm
e}^{-\frac{2}{3}\,{\frac {\sqrt {6} \kappa\,\phi}{\sqrt
{\alpha}}}}}-1\Big) \right]^{-2}\,.\hspace{0.5cm}
\end{eqnarray}
By obtaining $\phi_{hc}$ from equation (\ref{eq30}), substituting in
equation (\ref{eq35}) and considering that the expression ${\rm
e}^{-\frac{2}{3}\,{\frac {\sqrt {6}\kappa\,\phi_{hc}}{\sqrt
{\alpha}}}}$ is very small, we get
\begin{eqnarray}
\label{eq36}\epsilon=\frac{3}{4}\frac{\alpha}{N^{2}}\,,
\end{eqnarray}
which by using equation (\ref{eq17}) gives
\begin{eqnarray}
\label{eq37}r=\frac{12\alpha}{N^{2}}\,.
\end{eqnarray}
We see that, in large $N$ and small $\alpha$ limit, the
tensor-to-scalar ratio in DBI model is also the same as the
expression predicted for $r$ in the canonical single scalar field
model. The scalar spectral index in DBI model takes the following
form
\begin{eqnarray}
\label{eq38}n_{s}=1-\frac{18\alpha}{N^{2}}\left( \frac{\sqrt
{6}}{2N}\,\sqrt {{\frac {{\kappa}^{2}} {\alpha}}}\alpha+1 \right)
^{-2}+2\, \Bigg\{\frac{1}{2}\, \bigg[{\frac
{6V_{0}{\kappa}^{2}\alpha}{{N}^{2}}} -{\frac
{2V_{0}{\kappa}^{2}}{N}}-\frac{3}{2}\,{\frac
{V_{0}{\kappa}^{2}\alpha}{n{N}^{2}}}-2\, \bigg(\frac{3}{2}\,{\frac
{{ \kappa}^{2}\alpha}{{N}^{2}V_{0}}}+{\frac
{{\kappa}^{2}}{NV_{0}}}\nonumber\\
+\frac{3}{4}\,\frac{{\kappa} ^{2}\alpha}{V_{0}{n}{N}^{2}} \Big(
1-\frac{3}{4}\,{\frac {\alpha}{Nn}} \Big) ^{-2} \bigg) {V}^{2}_{0}
\bigg] {V}^{-1}_{0}-3/2\,{\frac {{\kappa}^{ 2}\alpha}{{N}^{2}}}
\Bigg\} {\kappa}^{-2}\,,\hspace{1cm}
\end{eqnarray}
where we have assumed $f_{0}^{-1}\equiv V_{0}$ for simplicity.
We note that although the functions $V(\phi)$ and
$f(\phi)$ are independent, however, both functions are E-model
(actually, the inverse of $f(\phi)$ is E-model). In the E-model
potential, the coefficient $V_{0}$ is an arbitrary constant. So,
when we adopt the E-model for the inverse of $f(\phi)$, the
coefficient $f_{0}$ also would be an arbitrary parameter. In this
regard, for simplicity, we adopt two constant as $f_{0}^{-1}\equiv
V_{0}$. The above scalar spectral index in the
large $N$ and small $\alpha$ limit becomes
\begin{eqnarray}
\label{eq39}n_{s}=1-\frac{4}{N}.
\end{eqnarray}
Here we see that in this limit, the scalar spectral index in DBI
model is somewhat different from the tachyon and canonical single
field models in the sense that the second term is $\frac{4}{N}$
(whereas in tachyon and canonical single field model is
$\frac{2}{N}$). In the $\alpha\rightarrow 0$ limit, $\epsilon$ tends to zero and deviation
of $n_s$ from the scale invariance comes from the value of $\eta$ in
this limit (see equation (\ref{eq16})). In a tachyon model (and also
canonical scalar field) $\eta$ is expressed in terms of the potential $V(\phi)$.
However, in DBI model, $\eta$ is function of both $V(\phi)$ and
$f(\phi)$ (see Eq. (\ref{eq29})) and both these functions contribute in deviation of the
scalar spectral index from unity. Considering that these two
functions in $\alpha\rightarrow 0$ limit are in the same order, the
deviation would be twice. The expression $n_s=1-\frac{4}{N}$ has
been obtained by the authors of Ref.~\cite{Li14} in a different manner.
By a field redefinition and adopting the quartic potential, they
obtained this expression for $c_s^2\ll 1$. However, in the current
work, we don't imply $c_s^2\ll 1$ limit. We obtain
$n_s=1-\frac{4}{N}$ by adopting E-model functions and considering
$\alpha\rightarrow 0$ limit.

Note that in $\alpha\rightarrow \infty$ limit, the E-model potential
tends to $\phi^{2n}$ and we have
\begin{equation}
\label{eq40}\epsilon={\frac {8\,{n}^{2}}{{\kappa}^{2} \left(
2\,n+\phi \right) ^{2}}}\,,
\end{equation}
and
\begin{eqnarray}
\label{eq41}n_{s}=1-\frac{48\,{n}^{2}}{{\kappa}^{2} \left( 2\,{\frac
{n}{\phi}}+1 \right) ^{2} {\phi}^{2}}-{\frac
{8n}{{\kappa}^{2}{\phi}^{2}}} \,.
\end{eqnarray}

We have performed a numerical study on the perturbation parameters
$r$ and $n_{s}$ and the results are shown in figure 3. In this
regard, we have used equations (\ref{eq16}) and (\ref{eq17}) with
the slow-roll parameters defined in equations (\ref{eq28}) and
(\ref{eq29}). As figure shows, the DBI model with E-model potential
in $\alpha\rightarrow \infty$ limit tends to the DBI model with
$\phi^{2n}$ potential. In $\alpha\rightarrow 0$ limit we have
$(n_{s}=0.92,r=0)$ for $N=50$ and ($n_{s}=0.933, r=0)$ for $N=60$.
The DBI model with E-model potential for both $n=1$ and $n=2$,
typically does not lie within the 95$\%$ confidence region of the
Planck2015 TT, TE, EE+lowP $r-n_{s}$ result. Nevertheless,
the values of the scalar spectral index in the DBI model with $n=1$
and $N=60$, in large $\alpha$ limit, are in $n_{s}=0.9652 \pm
0.0047$ range (this range is released by Planck2015 TT, TE, EE+ lowP
data).

\begin{figure}[ht]
\begin{center}
\includegraphics[scale=0.6]{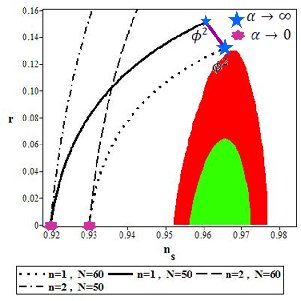}
\end{center}
\caption{\small {Tensor-to-scalar ratio versus the
spectral index for a DBI model with the E-model potential. The
smaller blue star is corresponding to DBI inflation with $\phi^{2n}$
potential and $N=50$ while the larger blue star is corresponding to
DBI inflation with $\phi^{2n}$ potential and $N=60$.}}
\label{fig3}
\end{figure}

\section{Reheating}\label{se3}

When the inflation phase terminates, the process of reheating take
places to reheat the universe for subsequent evolution. By studying
this process in the aforementioned models, we can find some
additional constraints on the model's parameter space. To this end,
we obtain some expressions for $N_{rh}$ and $T_{rh}$ (where
subscript \emph{rh} stands for reheating) in terms of the scalar
spectral index based on the strategy presented in
Refs.~\cite{Dai14,Un15,Co15,Cai15,Ue16}. The following expression
\begin{equation}
\label{eq42} N_{hc}=\ln \left(\frac{a_{e}}{a_{hc}}\right)\,,
\end{equation}
defines the e-folds number between the time of the horizon crossing
of the physical scales and the end of the inflationary expansion. In
this definition, $a_{e}$ is the scale factor at the end of the
inflation and $a_{hc}$ is the value of the scale factor at the
horizon crossing. During the reheating epoch we have the relation
$\rho\sim a^{-3(1+\omega_{eff})}$ for the energy density, in which
$\omega_{eff}$ is the effective equation of state of the dominant
energy density in the universe. In this respect, the e-folds number
of the reheating era in terms of the energy density and effective
equation of state is written as
\begin{eqnarray}\label{eq43}
N_{rh}=\ln\left(\frac{a_{rh}}{a_{e}}\right)=-\frac{1}{3(1+\omega_{eff})}\ln\left(\frac{\rho_{rh}}{\rho_{e}}\right)\,,
\end{eqnarray}
By setting the value of $k$ at horizon crossing by $k_{hc}$, we can
write
\begin{eqnarray}\label{eq44}
0=\ln\left(\frac{k_{hc}}{a_{hc}H_{hc}}\right)=
\ln\left(\frac{a_{e}}{a_{hc}}\frac{a_{rh}}{a_{e}}\frac{a_{0}}{a_{rh}}\frac{k_{hc}}{a_{0}H_{hc}}\right)\,,
\end{eqnarray}
where $a_{0}$ is the current value of the scale factor. From
equations (\ref{eq42}), (\ref{eq43}) and (\ref{eq44}) we obtain
\begin{eqnarray}\label{eq45}
N_{hc}+N_{rh}+\ln\left(\frac{k_{hc}}{a_{0}H_{hc}}\right)+\ln\left(\frac{a_{0}}{a_{rh}}\right)=0\,.
\end{eqnarray}

In the next step, it is useful to obtain an expression for
$\frac{a_{0}}{a}$ in terms of temperature and density. In this
regard, we use the following expression
\begin{equation}\label{eq46}
\rho_{rh}=\frac{\pi^{2}g_{rh}}{30}T_{rh}^{4}\,,
\end{equation}
which gives the relation between energy density and temperature in
reheating era \cite{Co15,Ue16}. The parameter $g_{rh}$ in equation
(\ref{eq46}) represents the effective number of the relativistic
species at the reheating epoch. On the other hand, from the
conservation of the entropy we have \cite{Co15,Ue16}
\begin{equation}\label{eq47}
\frac{a_{0}}{a_{rh}}=\left(\frac{43}{11g_{rh}}\right)^{-\frac{1}{3}}\frac{T_{rh}}{T_{0}}\,,
\end{equation}
where $T_{0}$ denotes the current temperature of the universe. By
using equation (\ref{eq46}) and (\ref{eq47}) we obtain the following
expression
\begin{eqnarray}\label{eq48}
\frac{a_{0}}{a_{rh}}=\left(\frac{43}{11g_{rh}}\right)^{-\frac{1}{3}}T_{0}^{-1}\left(\frac{\pi^{2}g_{rh}}{30\rho_{rh}}\right)^{-\frac{1}{4}}\,.
\end{eqnarray}

To proceed further and to obtain some explicit expressions for
$N_{rh}$ and $T_{rh}$, we should specify the model under
consideration. In this sense, in what follows we study non-canonical
tachyon and DBI models separately.

\subsection{Reheating in the tachyon model}\label{se3-1}
In a tachyon model, we can write the energy density in the following
form
\begin{eqnarray}\label{eq49}
\rho=\frac{V}{\sqrt{1-\frac{2}{3}\epsilon}}\,.
\end{eqnarray}
The energy density at the end of inflation era is obtained by
setting $\epsilon=1$ as follows
\begin{equation}\label{eq50}
\rho_{e}=\sqrt{3}\,V_{e}\,.
\end{equation}
Now, by using equations (\ref{eq43}) and (\ref{eq50}) we obtain
\begin{eqnarray}\label{eq51}
\rho_{rh}=\sqrt{3}\,V_{e}\exp\Big[-3N_{rh}(1+\omega_{eff})\Big].
\end{eqnarray}
From equations (\ref{eq48}) and (\ref{eq51}) we get
\begin{eqnarray}\label{eq52}
\ln\left(\frac{a_{0}}{a_{rh}}\right)=-\frac{1}{3}\ln\left(\frac{43}{11g_{rh}}\right)
-\frac{1}{4}\ln\left(\frac{\pi^{2}g_{rh}}{30\rho_{rh}}\right)-\ln
T_{0}
+\frac{1}{4}\ln\left(\sqrt{3}\,V_{e}\right)-\frac{3}{4}N_{rh}(1+\omega_{eff})\,.\hspace{1cm}
\end{eqnarray}
By using equation (\ref{eq12}), we can find $H_{hc}$. Then, from
equations (\ref{eq12}), (\ref{eq45}) and (\ref{eq52}), we obtain the
following expression for the e-folds number during reheating
\begin{eqnarray}\label{eq53}
N_{rh}=\frac{4}{1-3\omega_{eff}}\Bigg[-N_{hc}-\ln\Big(\frac{k_{hc}}{a_{0}T_{0}}\Big)-\frac{1}{4}\ln\Big(\frac{40}{\pi^{2}g_{rh}}\Big)
-\frac{1}{3}\ln\Big(\frac{11g_{rh}}{43}\Big)\nonumber\\+\frac{1}{2}\ln\Big(8\pi^{2}{\cal{A}}_{s}{\cal{W}}_{s}
c_{s}^{3}\Big) -\frac{1}{4}\ln\bigg(\sqrt{3}\, V_{e}\bigg)\Bigg].
\end{eqnarray}
The temperature during reheating is obtained from equations
(\ref{eq43}), (\ref{eq47}) and (\ref{eq50}) as follows
\begin{equation}\label{eq54}
T_{rh}=\bigg(\frac{30}{\pi^{2}g_{rh}}\bigg)^{\frac{1}{4}}\,
\bigg[\sqrt{3}\,
V_{e}\bigg]^{\frac{1}{4}}\,\exp\bigg[-\frac{3}{4}N_{rh}(1+\omega_{eff})\bigg]\,.
\end{equation}
To perform a numerical study, we should firstly rewrite equations
(\ref{eq53}) and (\ref{eq54}) in terms of the scalar spectral index.
In this regard, we use equation (\ref{eq1}) to rewrite equations
(\ref{eq53}) and (\ref{eq54}) in terms of the value of the scalar
field at horizon crossing ($\phi_{hc}$). Then, by considering that
$\phi_{hc}$ is related to $n_{s}$ (look at equations (\ref{eq1}),
(\ref{eq8}), (\ref{eq9}) and (\ref{eq16})), we can write $N_{rh}$
and $T_{rh}$ in terms of $n_{s}$ and then study the reheating phase
numerically. The results are shown in figures 4, 5 and 6. In figure
4, we have plotted the ranges of $N_{rh}$ and $\omega_{eff}$ which
lead to the observationally viable values of the scalar spectral
index. We have considered both $n=1$ and $n=2$ cases for
$\alpha=0.1$ and $\alpha \rightarrow \infty$. As figure 4 shows, in
all considered cases and with all assumed values of $\omega_{eff}$,
the instantaneous reheating (corresponding to $N_{rh}=0$, the point
in which all curves converge) is favored by Planck2015 observational
data, except for $n=2$ and $\alpha\rightarrow \infty$. The situation
is illustrated in figure 5 more explicitly. In this figure we have
plotted the e-folds number during reheating versus the scalar
spectral index for some sample values of the effective equation of
state. Figure 6 shows the temperature during reheating versus the
scalar spectral index.

We note that, in an inflation model with a
canonical scalar field, the e-folds number and temperature during
reheating are defined as equations (\ref{eq53}) and (\ref{eq54}).
However, the definitions of some parameters such as $N_{rh}$,
${\cal{A}}_{s}$ and ${\cal{W}}_{s}$ are different in the canonical
and non-canonical models. In the tachyon model, these parameters are
given by equations (\ref{eq11}), (\ref{eq12}) and (\ref{eq13}).
These parameters in a canonical model are defined as $ N\simeq
-\kappa^{2}\int_{\phi_{hc}}^{\phi_{e}} \frac{V }{V'} d\phi$, $
{\cal{A}}_{s}=\frac{H^{2}}{8\pi^{2}{\cal{W}}_{s}}$ and
${\cal{W}}_{s}=\frac {\dot{\phi}^{2}}{2H^{2}}$. These definitions
cause the different dependence of $N_{rh}$ and $T_{rh}$ to $\phi$
(or $\phi_{hc}$) and therefore to $n_{s}$. For instance, we have the
following expression in the canonical model~\cite{Ue16}
\begin{equation}\label{eqnew1}
N_{hc}=-\frac{3\alpha}{4n}\Bigg[e^{\sqrt{\frac{2\kappa^{2}}{3\alpha}}\phi_{e}}-e^{\sqrt{\frac{2\kappa^{2}}{3\alpha}}\phi_{hc}}
-\sqrt{\frac{2\kappa^{2}}{3\alpha}}\left(\phi_{e}-\phi_{hc}\right)\Bigg]\,.
\end{equation}
The corresponding parameter in the tachyon model is obtained as (see
equation (\ref{eq11}))
\begin{eqnarray}\label{eqnew2}
N_{hc}=\frac{3\kappa^{2}}{2}\,V_{0}\sqrt{\frac{3\alpha}{2\kappa^{2}}}\left(\phi_{e}-\phi_{hc}\right)-\frac{3}{8}V_{0}\alpha\left(
{\rm e}^{-2 \,{\sqrt {\frac{2\kappa^{2}}{3\alpha}}\,\phi_{e}}}- {\rm
e}^{-2 {\sqrt
{\frac{2\kappa^{2}}{3\alpha}}\,\phi_{hc}}}\right)\nonumber\\+\frac{9}{4}V_{0}\alpha\left({\rm
e}^{-{\sqrt {\frac{2\kappa^{2}}{3\alpha}}\,\phi_{e}}}-{\rm
e}^{-\sqrt {\frac{2\kappa^{2}}{3\alpha}}\,\phi_{hc}}\right)
-\frac{3}{4}V_{0}\alpha\left({\rm e}^{\sqrt
{\frac{2\kappa^{2}}{3\alpha}}\,\phi_{e}}-{\rm e}^{\sqrt
{\frac{2\kappa^{2}}{3\alpha}}\,\phi_{hc}}\right) \,.
\end{eqnarray}
As we can see, in the canonical model there are terms which are
linear and exponential in $\phi$. However, in the tachyon model,
there are also some terms which contain the inverse exponential of
$\phi$. Such expressions make the numerical results of two model
different. Let's consider the case with $n=1$ and $\alpha=0.1$. With
these choices and by adopting $\omega_{eff}=-\frac{1}{3}$, the
observational constraint on $N_{rh}$ for the canonical model is as
$N_{rh}\leq 4$ (see~\cite{Ue16}), while, the corresponding
constraint for the tachyon model is as  $N_{rh}\leq 26$. By adopting
$\omega_{eff}=0$, we have $N_{rh}\leq 8$ for the canonical
model~\cite{Ue16}, and $N_{rh}\leq 52$ for the tachyon model. These
mean that, in the non-canonical tachyon model, the reheating phase
can last longer than the reheating in the canonical model. We can
also compare the temperature during reheating in two models. For
$\omega_{ff}=-\frac{1}{3}$ in the canonical model, we have
$\log_{10}\left(\frac{T_{rh}}{GeV}\right)>14.3$~\cite{Ue16} and in
the tachyon model we have
$\log_{10}\left(\frac{T_{rh}}{GeV}\right)>0.9$. If we consider the
case with $\omega_{eff}=0$, for the canonical model we have
$\log_{10}\left(\frac{T_{rh}}{GeV}\right)>12.8$~\cite{Ue16} and for
the tachyon model there is no constraint on the temperature and for
any temperature we get the observationally viable $n_{s}$. Here
also, we see that in a non canonical tachyon model the larger range
of the temperature is corresponding to the observational viable
values of $n_{s}$.

\begin{figure}[ht]
\begin{center}
\includegraphics[scale=0.6]{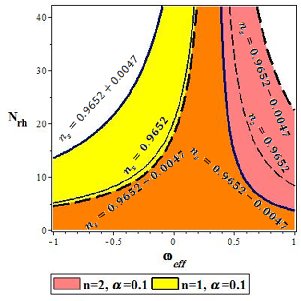}\includegraphics[scale=0.6]{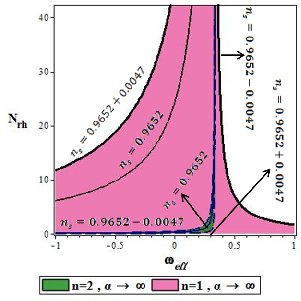}
\end{center}
\caption{\small {The ranges of the parameters $N_{rh}$
and $\omega_{eff}$ to have observationally viable values of the
scalar spectral index for a tachyon model with E-model potential.
The left panel corresponds to $\alpha$ =0.1 and the right one is for
$\alpha\rightarrow\infty$. Note that in the left panel, the yellow
region is bounded by solid lines and the red region is bounded by
the dashed lines. The orange overlap region is the range in which
both $ n=1$ and $n=2$ cases are consistent with observational data.
In the right panel, the magenta region is bounded by solid lines and
the green region is bounded with dashed lines and is actually the
overlap region in this case.}}
\label{fig4}
\end{figure}

\begin{figure}[ht]
\begin{center}
\includegraphics[scale=0.6]{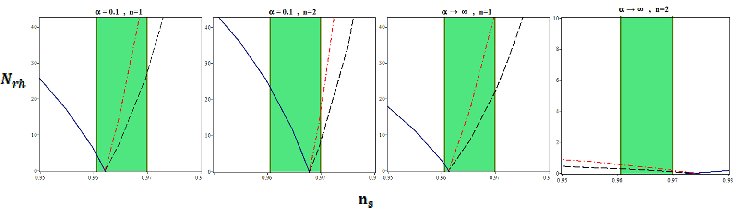}
\end{center}
\caption{\small {E-folds number during reheating epoch
versus the scalar spectral index in a tachyon model with E-model
potential. The dashed lines correspond to
$\omega_{eff}=-\frac{1}{3}$, the dashed-dotted lines correspond to
$\omega_{eff}=0$ and the solid lines correspond to $\omega_{eff}=1$.
The green region shows the values of $n_{s}$ released by Planck2015
experiment.}}
\label{fig5}
\end{figure}

\begin{figure}[ht]
\begin{center}
\includegraphics[scale=0.6]{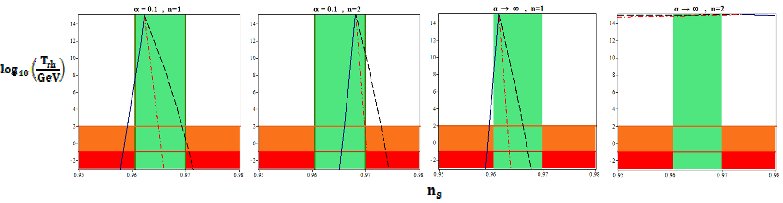}
\end{center}
\caption{\small {Temperature during reheating epoch
versus the scalar spectral index in a tachyon model with E-model
potential. The dashed lines correspond to
$\omega_{eff}=-\frac{1}{3}$, the dashed-dotted lines correspond to
$\omega_{eff}=0$ and the solid lines correspond to $\omega_{eff}=1$.
The orange region demonstrates the temperatures below the
electroweak scale, $T<100$ GeV and the red region shows the
temperatures below the big bang nucleosynthesis scale, $T<10$ MeV.}}
\label{fig6}
\end{figure}

\subsection{Reheating in the DBI model}\label{se3-2}
The energy density in the DBI model can be written as follows
\begin{eqnarray}\label{eq55}
\rho=\left(1+\frac{f^{-1}V^{-1}}{\sqrt{1+\frac{2}{3}\epsilon
(1+fV)}}\right) V\,.
\end{eqnarray}
By setting $\epsilon=1$, we obtain
\begin{equation}\label{eq56}
\rho_{e}=\Bigg(1+\frac{f_{e}^{-1}V_{e}^{-1}}{\sqrt{1+\frac{2}{3}(1+f_{e}V_{e})}}\Bigg)\,.
\end{equation}
The energy density during reheating era is obtained from equations
(\ref{eq43}) and (\ref{eq56}) as
\begin{eqnarray}\label{eq57}
\rho_{rh}=\Bigg(1+\frac{f_{e}^{-1}V_{e}^{-1}}{\sqrt{1+\frac{2}{3}(1+f_{e}V_{e})}}\Bigg)\,V_{e}\times
\exp\Big[-3N_{rh}(1+\omega_{eff})\Big].
\end{eqnarray}
Now, equations (\ref{eq48}) and (\ref{eq57}) give
\begin{eqnarray}\label{eq58}
\ln\left(\frac{a_{0}}{a_{rh}}\right)=-\frac{1}{3}\ln\left(\frac{43}{11g_{rh}}\right)
-\frac{1}{4}\ln\left(\frac{\pi^{2}g_{rh}}{30\rho_{rh}}\right)-\ln
T_{0}
+\frac{1}{4}\ln\left[\Bigg(1+\frac{f_{e}^{-1}V_{e}^{-1}}{\sqrt{1+\frac{2}{3}(1+f_{e}V_{e})}}\Bigg)
\,V_{e}\right]\nonumber\\-\frac{3}{4}N_{rh}(1+\omega_{eff})\,.\hspace{1cm}
\end{eqnarray}
From equations (\ref{eq12}), (\ref{eq45}) and (\ref{eq58}) we obtain
\begin{eqnarray}\label{eq59}
N_{rh}=\frac{4}{1-3\omega_{eff}}\Bigg[-N_{hc}-\ln\Big(\frac{k_{hc}}{a_{0}T_{0}}\Big)-\frac{1}{4}\ln\Big(\frac{40}{\pi^{2}g_{rh}}\Big)
-\frac{1}{3}\ln\Big(\frac{11g_{rh}}{43}\Big)+\frac{1}{2}\ln\Big(8\pi^{2}{\cal{A}}_{s}{\cal{W}}_{s}
c_{s}^{3}\Big)
\nonumber\\-\frac{1}{4}\ln\bigg(\bigg(1+\frac{f_{e}^{-1}V_{e}^{-1}}{\sqrt{1+\frac{2}{3}(1+f_{e}V_{e})}}\bigg)\,
V_{e}\bigg)\Bigg].\hspace{1cm}
\end{eqnarray}
Also, from equations (\ref{eq43}), (\ref{eq47}) and (\ref{eq56}) we
get
\begin{eqnarray}\label{eq60}
T_{rh}=\bigg(\frac{30}{\pi^{2}g_{rh}}\bigg)^{\frac{1}{4}}
\left[\Bigg(1+\frac{f_{e}^{-1}V_{e}^{-1}}{\sqrt{1+\frac{2}{3}(1+f_{e}V_{e})}}\Bigg)\,
V_{e}\right]^{\frac{1}{4}} \times
\exp\bigg[-\frac{3}{4}N_{rh}(1+\omega_{eff})\bigg]\,.\hspace{1cm}
\end{eqnarray}

By rewriting the equations (\ref{eq59}) and (\ref{eq60}) in terms of
the scalar spectral index (similar to what we have done in the
tachyon model), we can perform numerical analysis in this model.
Note that since the DBI model with E-model potential for $n=2$ is
not consistent with the observational data, we don't study reheating
in this case. However, in the $n=1$ case, the scalar spectral index
is consistent with observation (although $r$ is not), so we explore
reheating in this case. Actually, the observationally
viable values of the scalar spectral index can set some constraints
on the reheating parameters in DBI model. We remember, for instance, that in
a two-field inflation model, one field is responsible for inflation
and reheating and the other one is important in perturbations. If
we consider DBI as a field responsible for inflation and reheating and not for perturbations
in a two-field model, the value of the
tensor-to-scalar ratio no matters. In this regard, we think it
makes sense to explore the reheating phase for DBI model to see its
cosmological consequences. The results are shown
in figures 7, 8 and 9. In figure 7 we have plotted the region of the
e-folds number during reheating and the effective equation of state
for which the scalar spectral index in a DBI model with E-model
potential (for $n=1$) is consistent with Planck2015 observational
data. As this figure shows, with $\alpha=0.1$ and $n=1$, the
instantaneous reheating is disfavored by Planck2015 data for all
values of $\omega_{eff}$ (between $-1$ and $+1$). However, with
$\alpha\rightarrow\infty$ and $n=1$, for all values of the effective
equation of state parameter (varying between $-1$ and $+1$) the
instantaneous reheating is favored by observational data.
In fact, these results confirm the ones obtained in
section \ref{se2-2}, in the sense that the scalar spectral index
(and therefore the e-folds number and temperature during reheating)
in large $\alpha$ limit is observationally viable.
These situations are clarified also in figure 8.
In figure 9 we have plotted the temperature during reheating versus
the scalar spectral index.

\begin{figure}[ht]
\begin{center}
\includegraphics[scale=0.6]{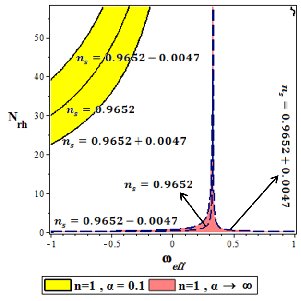}
\end{center}
\caption{\small {The ranges of the parameters $N_{rh}$
and $\omega_{eff}$ which lead to the observationally viable values
of the scalar spectral index for a DBI model with E-model
potential.}}
\label{fig7}
\end{figure}

\begin{figure}[ht]
\begin{center}
\includegraphics[scale=0.6]{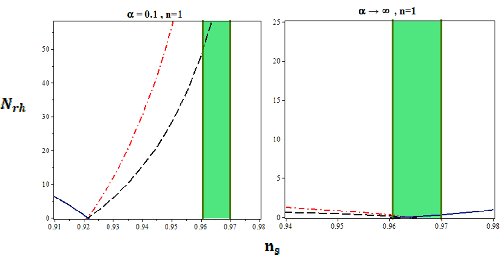}
\end{center}
\caption{\small {E-folds number during reheating versus
the scalar spectral index in a DBI model with E-model potential. The
dashed lines correspond to $\omega_{eff}=-\frac{1}{3}$, the
dashed-dotted lines correspond to $\omega_{eff}=0$ and the solid
lines correspond to $\omega_{eff}=1$. The green region shows the
values of $n_{s}$ released by the Planck2015 dataset.}}
\label{fig8}
\end{figure}

\begin{figure}[ht]
\begin{center}
\includegraphics[scale=0.6]{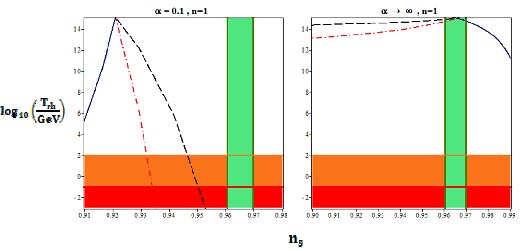}
\end{center}
\caption{\small {Temperature during reheating versus the
scalar spectral index in a DBI model with E-model potential. The
dashed lines correspond to $\omega_{eff}=-\frac{1}{3}$, the
dashed-dotted lines correspond to $\omega_{eff}=0$ and the solid
lines correspond to $\omega_{eff}=1$. The orange region demonstrates
the temperatures below the electroweak scale, $T<100$ GeV and the
red region shows the temperatures below the big bang nucleosynthesis
scale, $T<10$ MeV.}}
\label{fig9}
\end{figure}

Note that, with $\omega_{eff}=-\frac{1}{3}$, by repeating the
analysis performed to obtain equations (\ref{eq53}), (\ref{eq54}),
(\ref{eq59}) and (\ref{eq60}) we cannot obtain analytical closed
expressions for number of e-folds and temperature. However, a
vertical line in the plots can be a curve for
$\omega_{eff}=-\frac{1}{3}$ which crosses the instantaneous
reheating point~\cite{Un15,Co15}.

\section{Summary and Discussion}\label{se4}

In this paper, we have considered two non-canonical scalar field
models: tachyon and DBI models. Motivated by the $\alpha$-attractor
models, we have adopted the E-model potential to seek for
$\alpha$-attractor in these models. We have calculated the slow-roll
parameters, scalar spectral index and tensor-to-scalar ratio in both
models. The tachyon model with E-model potential in large $N$ and
small $\alpha$ limit predicts the value of the scalar spectral index
and tensor-to scalar ratio as $n_{s}=1-\frac{2}{N}$ and
$r=\frac{12\alpha}{N^{2}}$. These predicted parameters are exactly
the same as the ones predicted in the canonical single field model
with E-model potential. In $\alpha \rightarrow \infty$ limit, the
tachyon model with E-model potential reaches the model with
$\phi^{2n}$ potential. We have also analyzed the tachyon model
numerically and compared the results with the Planck2015 TT, TE,
EE+lowP observational data. We have found that the tachyon model
with E-model potential and with both $N=50$ and $N=60$ for all
values of $\alpha$ is consistent with the observational data. The
$r-n_{s}$ trajectories with a given value of the e-folds number, for
both $n=1$ and $n=2$ reaches a fixed point. This means that for
$\alpha\rightarrow 0$ the values of the scalar spectral index and
tensor-to-scalar ratio are independent of $n$. The value of the
scalar spectral index and tensor-to-scalar ratio in small $\alpha$
limit, predicted by DBI model, are as $n_{s}=1-\frac{4}{N}$ and
$r=\frac{12\alpha}{N^{2}}$. In DBI model, the calculated $r$ is the
same as the one predicted in tachyon and canonical scalar field
models. However, $n_{s}$ is somewhat different in the sense that the
second term is $\frac{4}{N}$, a factor of 2 different with the
corresponding term in tachyon case. Numerical analysis of the DBI
model and comparing with the observational data shows that the DBI
model with E-model potential does not lie within the 95$\%$
confidence region of the $n_{s}-r$ plane released by Planck2015.
But, in large $\alpha$ limit, the value of the scalar spectral index
is consistent with observation, though the value of the
tensor-to-scalar ratio is not. For $N=50$, the value of $n_{s}$ in
the DBI model with $\alpha\rightarrow\infty$ is consistent with
Planck2015 TT, TE, EE+lowP observational data. For $N=60$, the value
of $n_{s}$ in the DBI model with $\alpha>10^{4}$ is consistent with
the observational data.

The reheating era after inflation epoch also has been studied in
this paper. For both treated models, we have obtained some
expressions for the e-folds number and temperature during the
reheating era which give some additional constraints on the model's
parameters space. We have studied the parameters $N_{rh}$, $T_{rh}$
and $\omega_{eff}$ numerically and the results have been shown in
figures. By considering the values of the scalar spectral index,
allowed by Planck2015 TT, TE, EE+lowP data, we have plotted the
regions of $N_{rh}$ and $\omega_{eff}$ which are observationally
viable. For tachyon model, we have adopted both $n=1$ and $n=2$ with
both $\alpha=0.1$ and $\alpha\rightarrow\infty$. Our numerical
analysis shows that, for $n=1$ with both $\alpha=0.1$ and
$\alpha\rightarrow\infty$ and for $n=2$ with $\alpha=0.1$, the
instantaneous reheating is favored by Planck2015 data. For $n=2$ and
$\alpha\rightarrow\infty$, the instantaneous reheating is disfavored
by the observational data. We have obtained some constraints by
adopting these sample values of the parameters. The constraints on
the tachyon model's parameters, obtained by studying $N_{rh}$ and
$n_{s}$ are summarized in table 1.

\begin{small}
\begin{table*}
\caption{\label{tab:2} The ranges of the number of e-folds parameter
and temperature for tachyon model at reheating which are consistent with observational
data.}
\begin{tabular}{cccccccc}
\\ \hline \hline&$n=1\,\,,\,\,\alpha=0.1$&&$n=1\,\,,\,\,\alpha\rightarrow\infty$&&
$n=2\,\,,\,\,\alpha=0.1$&&$n=2\,\,,\,\,\alpha\rightarrow\infty$\\ \hline\\
$\omega_{eff}=-\frac{1}{3}$&  $N_{rh}<26$ &&$N_{rh}<21$&& $N_{rh}<8$
&&$0.09<N_{rh}<0.026$\\\\$\omega_{eff}=0$& $N_{rh}<52$
&&$N_{rh}<42$&& $N_{rh}<15.5$
&&$0.2<N_{rh}<0.6$\\\\$\omega_{eff}=1$& $N_{rh}<4.5$
&&$N_{rh}<1.9$&& $N_{rh}<23$
&&--------\\\\
\hline $\omega_{eff}=-\frac{1}{3}$&
$\log_{10}\left(\frac{T_{rh}}{GeV}\right)>0.9$ &&--------&&
$\log_{10}\left(\frac{T_{rh}}{GeV}\right)>10.6$
&&$14.97>\log_{10}\left(\frac{T_{rh}}{GeV}\right)>14.93$\\\\$\omega_{eff}=0$&
--------
&&--------&& $\log_{10}\left(\frac{T_{rh}}{GeV}\right)>1.8$
&&--------\\\\$\omega_{eff}=1$&
$\log_{10}\left(\frac{T_{rh}}{GeV}\right)>7.4$
&&$\log_{10}\left(\frac{T_{rh}}{GeV}\right)>8.1$&&
--------
&&$\log_{10}\left(\frac{T_{rh}}{GeV}\right)>8.1$\\\\
\hline
\end{tabular}
\end{table*}
\end{small}

Studying the temperature during reheating era gives some more
constraints. The constraints, which are based on the observationally
viable values of the scalar spectral index, are presented in
table 1.

Regarding that the DBI model with E-model potential and with $n=2$
is not consistent with the observational data, we have performed the
numerical analysis on the reheating issue with $n=1$. The numerical
study shows that in this model with $n=1$ and $\alpha=0.1$, the
instantaneous reheating is disfavored by Planck2015 data (note that,
the scalar spectral index also in the case with $n=1$ and
$\alpha=0.1$ is disfavored by observational data). However, with
$n=1$ and $\alpha\rightarrow\infty$ the instantaneous reheating is
favored by the observation. Studying $N_{rh}$ and $T_{rh}$ gives
also some more constraints based on the viable values of $n_{s}$,
which are summarized in table 2.

\begin{small}
\begin{table*}
\caption{\label{tab:2} The ranges of the number of e-folds parameter
and temperature for DBI model at reheating which are consistent with observational
data.}
\begin{tabular}{cccccccccc}
\\ \hline \hline&$n=1\,\,,\,\,\alpha=0.1$&&$n=1\,\,,\,\,\alpha\rightarrow\infty$&&
&&&&$n=1\,\,,\,\,\alpha\rightarrow\infty$\\ \hline\\
$\omega_{eff}=-\frac{1}{3}$&  $49<N_{rh}<80$ &&$N_{rh}<0.12$&&&&
 &&$\log_{10}\left(\frac{T_{rh}}{GeV}\right)>14.9$\\\\$\omega_{eff}=0$& $100<N_{rh}<160$
&&$N_{rh}<0.25$&&
&&&&$\log_{10}\left(\frac{T_{rh}}{GeV}\right)>14.70$\\\\$\omega_{eff}=1$&
-------- &&$N_{rh}<0.31$&&
&&&&$\log_{10}\left(\frac{T_{rh}}{GeV}\right)>14.77$\\\\
\hline
\end{tabular}
\end{table*}
\end{small}

For the case with $n=1$ and $\alpha=0.1$, there is no constraint on
the reheating temperature.

It seems that if we consider a non-canonical scalar field with the
E-model type of potential, the tachyon model is more consistent with
observational data than the DBI model. In the tachyon model, the
values of the scalar spectral index and tensor-to-scalar ratio for
all values of $\alpha$ are consistent with Planck2015 data. Also,
there is an attractor point in this model which its scalar spectral
index is observationally viable. Exploring the reheating era in this
model shows also that this model is observationally viable.

Finally, we note that it would be interesting to think about if
one consider some kinetic driven models, like
k-inflation~\cite{Arm99}, and consider the nonminimal coupling term
and potential to be E-model. In this case also, we probably get
similar attractors. This is because the E-model function and
potential in the small $\alpha$ limit tend to a constant and so we
probably get some attractors in this limit. \\

{\bf Acknowledgement}\\
We would like to thank the referee for very insightful comments that
improved the quality of the paper considerably.
This work has been supported financially by Research
Institute for Astronomy and Astrophysics of Maragha (RIAAM) under
research project number 1/5237-**.

\end{document}